\documentclass[fleqn,12pt]{article}
\usepackage{epsfig}
\begin{document}
\textheight 22cm
\textwidth 15cm
\noindent
{\Large \bf Non-perturbative models of intermittency in edge turbulence}
\newline
\newline
Johan Anderson\footnote{anderson.johan@gmail.com} and Eun-jin Kim
\newline
University of Sheffield
\newline
Department of Applied Mathematics
\newline
Hicks Building, Hounsfield Road 
\newline 
Sheffield
\newline
S3 7RH
\newline
UK
\newline
\newline
\begin{abstract}
\noindent
A theory of the probability distribution function (PDF) tails of the blob density in plasma edge turbulence is provided. A simplified model of the fast convective radial transport is used. The theoretically predicted PDF tails corroborate earlier measurements of edge transport, further confirming the strongly non-Gaussian feature of edge transport. It is found that increasing the cross sectional spatial scale length ($L_x$ and $L_y$) of the blob results in larger transport whereas increasing the toroidal scale length ($L_z$) decreases the PDF. The results imply that the PDF decreases for larger blob speed $v_b$.
\end{abstract}
\newpage
\renewcommand{\thesection}{\Roman{section}}
\section{Introduction}
It is well known that turbulent transport determines the confinement of plasmas in magnetic fusion devices. These inherently nonlinear phenomena are rather complex and still not well understood since they vary from improved confinement regimes to very violent disruptions. One important observation from experiments is that overall improved confinement is found when the edge turbulence is suppressed. Edge plasma turbulence is crucial for wall erosion and plasma contamination through the transport of particles and heat to the vessel walls, and thus for the confinement in future reactors~\cite{a10}-~\cite{a11}. Interestingly, experimental measurements of edge turbulence has shown the highly intermittent nature~\cite{a51}-~\cite{a53}. Furthermore simulations of statistical properties of edge turbulence in 2D~\cite{a31}-~\cite{a33} and 3D~\cite{a41}-~\cite{a43} have revealed generic non-Gaussian probability distribution functions (PDF) of fluctuation levels. In particular, in the turbulence simulation a large fraction of large events (or blobs) that ballistically propagate in the radial direction have been observed~\cite{a15}-~\cite{a18}, which are especially dangerous for confinement.   
 
The so-called blob is a coherent structure with a higher density than the surrounding plasma which is localized in a plane perpendicular to the magnetic field $\vec{B}$ while extended along the field line. When a charge dependent drift such as those induced by curvature or centrifugal force is present, the blob becomes polarized as the effective sheath resistivity creates an electric field. The resulting $\vec{E} \times \vec{B}$ drift transports the blob to the outer wall. The natural outward convective transport of blobs in edge plasmas indicates that these coherent structures may play a crucial role in intermittency and the non-Gaussian statistics in edge plasmas~\cite{a11}.

Coherent structures such as blobs, streamers or vortices are often associated with avalanche like events of large amplitude and can therefore be of great importance for transport dynamics. Although these events are relatively rare, they can carry more than 50\% of the total fluxes~\cite{a60}-~\cite{a61}. Conventional methods to characterize transport have been limited to mean field theory, where the transport is described by one averaged coefficient. There are however at present a lot of evidence that transport often involves events of many different amplitudes or scales, some of which are intermittent and bursty in time. Since these intermittent and bursty events are highly non-linear phenomena contributing to the non-Gaussian structure of the PDF tails, they are poorly described by mean field theory. To characterize the intermittent turbulent transport a non-perturbative way is needed~\cite{a19}-~\cite{a23}.

In this paper we present a non-perturbative analytical theory of the PDF tails of density blob formation in tokamak edge plasmas. By adopting a simple nonlinear fluid theory of the blobs~\cite{a12}-~\cite{a14}, we predict that the blob density PDF tails have the exponential dependency ($P(n_b) \sim \exp{\{ -\zeta n_b^3\}}$), where $n_b$ is the blob density and $\zeta$ is a coefficient dependent on the blob properties. Note that this scaling is similar to what was found for zonal flow structure formation in ion-temperature-gradient mode turbulence~\cite{a23}. Interestingly, this exponential scaling agrees rather well with previous experimental results with reasonable values of the coefficient $\zeta$ for parameter values typical of plasma blobs~\cite{a11}.

Furthermore, we have found that increasing the cross sectional spatial scale length ($L_x$ and $L_y$) of the blob results in larger transport whereas increasing the toroidal scale length ($L_z$) decreases the PDF. Interpreting the constants in the blob speed seems to indicate that the PDF decreases for larger $v_b$. Note however that the blob speed is not a fundamental parameter in our model, but rather a combination of other parameters.

The paper is organized as follows. In Sec. II the physical model of the blob density is presented together with preliminaries of the path-integral formulation for the PDF tails of structure formation. In Sec III the instanton solutions are calculated and in Sec IV the PDF tails of blob formation are estimated. We provide numerical results in Sec. V and a discussion of the results and conclusion in Sec. VI. 

\section{Non-perturbative calculation of structure formation PDF}
The derivation of the physical model for radial plasma transport closely follows Ref.~\cite{a12}. We assume that the Scrape-off-Layer (SOL) plasma temperature, $T$, is constant. The electrostatic potential $\phi$ is constant along the magnetic field $\vec{B}$ and can be calculated from the equation for electric current
\begin{eqnarray}
\nabla \vec{j}_{\perp} + \nabla_{\parallel} \vec{j}_{\parallel} = 0,
\end{eqnarray}
with $j_{\perp} = c (\vec{B} \times \nabla P)/B^2$, where $P = nT$, $n$ is the plasma density and $c$ is the speed of light. Performing an integration along the field line and using the boundary conditions $j_{\parallel}|_{target} = e n_t c_s \frac{e \phi}{T}$ at the targets we find
\begin{eqnarray}
\frac{e \phi}{T} = \frac{\rho_i}{2 n_t B} \int dl \nabla \ln B \cdot(\vec{B} \times \nabla n)
\end{eqnarray}
where we have assumed that $|\frac{e \phi}{T}|< 1$; $n_t$ is the plasma density at the targets, $c_s = \sqrt{T/M}$ is the sound speed, $M$ is the ion mass, $e$ is the electron charge, $\rho_i$ is the ion gyro-radius, and the coordinate $l$ goes along the magnetic field line. For a plasma blob with density $n_b$ with parallel length $L_z$ situated around the midplane, Eq. (2) gives
\begin{eqnarray}
\frac{e \phi}{T} = \frac{L_z \rho_i}{2 R n_t} \frac{\partial n_b}{\partial y},
\end{eqnarray}
where we have neglected the magnetic shear and used $\nabla \ln B = e_x / R$; $R$ is the major radius and $x$ and $y$ are the local coordinates along the radial and poloidal directions, respectively. By using Eq. (3) for the $\vec{E} \times \vec{B}$ drift velocity, we find the blob plasma continuity equation in the form
\begin{eqnarray}
\frac{\partial n_b}{\partial t} + \frac{c_s \rho_i^2 L_z}{2R} \left( \frac{\partial}{\partial x} [n_b \frac{\partial }{\partial y}(\frac{1}{n_t} \frac{\partial n_b}{\partial y})] - \frac{\partial}{\partial y} [n_b \frac{\partial }{\partial x}(\frac{1}{n_t} \frac{\partial n_b}{\partial y}) \right) = f 
\end{eqnarray}
When $n_t = \xi n_b$ for a constant $\xi$, the separation of variables gives
\begin{eqnarray}
n_b(t,x,y) = n_0 (x,t) e^{-(y/L_y)^2},
\end{eqnarray}
reducing Eq. 4 to a ballistic equation for $n_0$
\begin{eqnarray}
(\frac{\partial}{\partial t} + v_b \frac{\partial }{ \partial x}) n_0 (t,x) = 0,
\end{eqnarray}
with
\begin{eqnarray}
v_b = c_s \left(\frac{\rho_i}{L_y}\right)^2 \frac{L_z}{R} \frac{n_b}{n_t}.
\end{eqnarray}
Note that the separable solution does not set the radial scale of the blob. The forcing $f$ is defined in Eq. 8. 

There has been suggestion from both simulations and experiments is that the blob is formed from the non-linear saturation of the linear instabilities at the plasma edge~\cite{a11}. Note that in the formation zone an approximately equal amount of enhanced density blobs and holes are generated~\cite{a11}. The effective gravity (polarization) causes these newly formed coherent structures to move, the blobs move outwards whereas the holes move inwards. The ballistic equation describing the dynamics is symmetric under the change of parameters ($x \rightarrow -x$, $n_0 \rightarrow -n_0$)~\cite{a81}. Note also that the blob velocity $v_b$ in Eq. (7) changes sign under this transformation. A detailed mechanism for the source of blobs is outside the scope of the present paper. In the following, we thus simply assume that there is a stochastic forcing (e.g. due to instabilities) and investigate the likelihood [probability distribution function (PDF)] of blob formation triggered by this forcing. Due to the stochastic forcing, blobs become short-lived in time, as shall be seen later. 

In order to calculate the PDF tails of blob formation, we utilize the instanton method~\cite{a25}. To this end, the PDF tail is expressed in terms of a path integral by utilizing the Gaussian statistics of the forcing $f$~\cite{a25}. We assume the statistics of the forcing $f$ to be Gaussian with a short correlation time modeled by the delta function as
\begin{eqnarray}
\langle f(x, t) f(x^{\prime}, t^{\prime}) \rangle = \delta(t-t^{\prime})\kappa(x-x^{\prime}),
\end{eqnarray}
and $\langle f \rangle = 0$.
The delta correlation in time was chosen for the simplicity of the analysis. In the case of a finite correlation time the non-local integral equations in time are needed. Note that the forcing $f$ was chosen to excite blobs; the source of the forcing is assumed to be the fluctuations due to instability. The spatial overlap between the forcing and the blob is critical for the generation of a blob. 

The probability distribution function of blob density $n_b$ can be defined as
\begin{eqnarray}
P(Z) & = &  \langle \delta(n_b - Z) \rangle \nonumber \\
& = & \int d \lambda  \exp(i \lambda Z) \langle \exp(-i \lambda n_b) \rangle \nonumber \\
& = & \int d\lambda \exp(i \lambda Z) I_{\lambda},
\end{eqnarray}
where 
\begin{eqnarray}
I_{\lambda} = \langle \exp(-i \lambda n_b) \rangle.
\end{eqnarray}
The angular brackets denote the average over the statistics of the forcing $f$. The integrand can then be rewritten in the form of a path-integral as
\begin{eqnarray}
I_{\lambda} = \int \mathcal{D} n_b \mathcal{D} \bar{n}_b e^{-S_{\lambda}}.
\end{eqnarray}
where
\begin{eqnarray}
S_{\lambda} & = & -i \int d^2x dt \bar{n}_b \left( \frac{\partial n_b}{\partial t} + \frac{c_s \rho_i^2 L_z}{2R} \left( \frac{\partial}{\partial x} [n_b \frac{\partial }{\partial y}(\frac{1}{n_t} \frac{\partial n_b}{\partial y})] - \frac{\partial}{\partial y} [n_b \frac{\partial }{\partial x}(\frac{1}{n_t} \frac{\partial n_b}{\partial y}) \right) \right) \nonumber \\
& + & \frac{1}{2} \int d^2x d^2 x^{\prime} dt  \bar{n}_b(x,t) \kappa(x-x^{\prime}) \bar{n}_b(x^{\prime},t) \nonumber \\
& + & i \lambda \int d^2x dt n_{b}(t) \delta(t).
\end{eqnarray}
Note that $P(Z)$ represents the probability of blob density taking a value $Z$. 

Note that the PDF tails of blob density can be found by calculating the value of $S_{\lambda}$ at the saddle-point in the case $\lambda \rightarrow \infty$. This will be done in Sec. III - IV.

\section{Instanton (saddle-point) solutions}
We have now reformulated the problem of calculating the PDF to a path-integral in Eq. (9). Although the path integral cannot in general be calculated exactly, an approximate value can be found in the limit $\lambda \rightarrow \infty$ by using a saddle point method to compute PDF tail. Since a direct application of the saddle-point equations results in very complicated partial differential equations for $n_b$ and $\bar{n}_b$, we assume that the instanton saddle-point solution is a temporally localized blob. That is, we assume that a short lived non-linear blob solution exists to the system of Eq. (4) in the form of a ballistically traveling solution Eqs (5) and (6). The blob density instanton takes the form $n_b(x,y,t) = n_0(x,y,t)F(t)$ while the target density is assumed to be $n_t = \xi n_0$. Here $n_0(x,y,t)= n_0(x-v_b t) e^{-(y/L_y)^2}$ denotes the spatial form of the coherent structure or blob and $F(t)$ is a temporally localized amplitude, representing the creation process.

The action $S_{\lambda}$ consists of three different parts; the blob model, the forcing and structure formation, respectively. The full action including the forcing and structure formation terms can then be expressed in terms of the time dependent function $F$ and the conjugate variable $\bar{N}$,
\begin{eqnarray}
S_{\lambda} & = & -i \int dt \bar{N} \left( \dot{F} + \frac{v_b}{L_x} F + K \frac{2}{L_y^2 L_x \xi} F^2 \right) \nonumber \\
& + & \frac{1}{2} \kappa_0 \int dt \bar{N}^2 \nonumber \\
& + &  i \lambda N \int dt F  \delta(t).
\end{eqnarray}
Here,
\begin{eqnarray}
N & = & \int d^2 x n_0(x-v_b t) e^{-(y/L_y)^2}, \\
\bar{N} & = & \int d^2 x \bar{n}_b(t, x) n_0(x-v_b t) e^{-(y/L_y)^2}, \\
K & = & \frac{c_s \rho_i^2 L_z}{2R},
\end{eqnarray}
and the radial scale-length ($L_x$) is defined as
\begin{eqnarray}
\frac{\partial n_0}{\partial x} = - \frac{1}{L_x} n_0. 
\end{eqnarray}
$\kappa_0$ in Eq. (13) is the strength of the forcing function $\kappa(x-x^{\prime})$, which is approximated by Taylor expansion in $x$ and $x^{\prime}$ for simplicity. Keeping only the zeroth order terms in the expansion gives us the separable integral in $x$ and $x^{\prime}$. {The time dependent function $N$ is the mean value averaged over the blob, $\bar{N}$ is the conjugate variable acting as a mediator between the real variable ($N$) and the forcing ($f$) and $K$ is a constant used to simplify the expressions.}

The saddle point equations for instantons (the equations of motion) are obtained by minimizing the effective action $S_{\lambda}$ with respect to the independent variables $F$ and $\bar{N}$:
\begin{eqnarray}
\frac{\delta S_{\lambda}}{\delta \bar{N}} & = & -i\left( \dot{F} + \frac{v_b}{L_x} F + K \frac{2}{L_y^2 L_x \xi}F^2 \right) + \kappa_0 \bar{N} = 0, \\
\frac{\delta S_{\lambda}}{\delta F} & = & -i\left( - \dot{\bar{N}} + \frac{v_b}{L_x} \bar{N} + K \frac{2}{L_y^2 L_x \xi}F \bar{N}\right) - \lambda N \delta(t) = 0.
\end{eqnarray}
The equation of motion for $F$ is derived for $t < 0$ using Eqs. (18)-(19) as,
\begin{eqnarray}
\frac{1}{2} \frac{d \dot{F}^2}{dF}& = & \eta_1 F + 6 \eta_2 F^2 + 8 \eta_3 F^3,
\end{eqnarray}
where
\begin{eqnarray}
\eta_1 & = & (\frac{v_b}{L_x})^2, \\
\eta_2 & = & \frac{v_b K}{L_y^2 L_x^2 \xi} = \sqrt{\eta_1 \eta_3}, \\
\eta_3 & = &  \frac{K^2}{L_y^4 L_x^2 \xi^2}. 
\end{eqnarray}
Note that the constants $\eta_1$, $\eta_2$ and $\eta_3$ all have the dimension of $1/[\mbox{Time}]^2$. We first perform an integration in $F$ and then use separation of variables to find
\begin{eqnarray}
\int \frac{dF}{F \sqrt{\eta_1 + 4 \eta_2 F + 4 \eta_3 F^2}} & = & \frac{ 1}{2 \sqrt{\eta_3}} \int \frac{dF}{F (F + \frac{\eta_2}{2 \eta_3})} \nonumber \\
 \frac{\sqrt{\eta_3}}{\eta_2} \ln (\frac{F}{F + \frac{\eta_2}{2 \eta_3}} \times  \frac{F_0 + \frac{\eta_2}{2 \eta_3}}{F_0} )& = & \int dt = t.
\end{eqnarray}
We then solve for $F$ to find
\begin{eqnarray}
H(t) & = & \frac{F_0}{F_0 + \frac{\eta_2}{2 \eta_3}}\exp{\{ \frac{\eta_2}{\sqrt{\eta_3}}t\}}, \\
F(t) & = & \frac{\eta_2}{2 \eta_3} \frac{H(t)}{1 - H(t)}.
\end{eqnarray}
Note that $H(t)<1$ and in Eq. (26) we impose the boundary conditions $F(t \rightarrow - \infty) \rightarrow 0$ and $F(0) = F_0$. The initial condition for $F(0) = F_0$ is found by integrating Eq. (19) over the time interval $(-\epsilon, \epsilon)$
\begin{eqnarray}
i (\bar{N}(\epsilon) - \bar{N}(-\epsilon)) = \lambda N,
\end{eqnarray}
where we use the fact that the conjugate field $\bar{N}$ disappears for positive time ($t>0$), which gives a relationship between the conjugate field $\bar{N}$ and the large factor $\lambda$
\begin{eqnarray}
\bar{N}(-\epsilon) = i \lambda N.
\end{eqnarray}
The property of the conjugate variables to vanish for ($t>0$) can be interpreted as a causality condition. Using Eq. (18) at $t=0$, we obtain
\begin{eqnarray}
-i \left( \dot{F} +\frac{v_b}{L_x} F_0 + \frac{2}{L_y^2 L_x \xi} K F_0^2\right) = -\kappa_0 \bar{N} = -i \lambda \kappa_0 N.
\end{eqnarray}
In the large $\lambda$ limit this gives
\begin{eqnarray}
F_0 = \pm \frac{1}{2\sqrt{K}}\sqrt{L_y^2 L_x \xi \kappa_0 N} \sqrt{\lambda}.
\end{eqnarray}
The instanton solution is localized within a time interval proportional to $1/\sqrt{\eta_1}$. The time scale of the blob or instanton solution can be estimated as $\tau_{blob} = \eta_1 = \frac{L_x}{v_b} = 10^{-4}s$, by using values in Ref.~\cite{a11}. Comparing this blob time scale with that of the ambient fluctuations ($\tau_{turb} = 10^{-5}s$) we find that $\tau_{turb}<\tau_{blob}$. This suggests that blobs described by this model with lifetime longer than the turbulence may significantly contribute to intermittent phenomena in the tokamak edge.

\section{The PDF tails}
We will now compute $\lambda$ dependence of $S_{\lambda}$ in Eq (13) at the saddle point i.e. the saddle point action which will then determine the PDF
\begin{eqnarray}
S_{\lambda} & = & -i \int dt \bar{N} \left( \dot{F} + \frac{v_b}{L_x} F + \gamma F^2 \right) \nonumber \\
& + & i \lambda N \int dt F \delta(t) + i \frac{\kappa_0}{2} \int dt \bar{N}^2 \nonumber \\
& = & \frac{1}{2 \kappa_0} \int dt \left(\dot{F} + \frac{v_b}{L_x} F + \gamma F^2 \right)^2 + i \lambda N F_0 \\
& \approx & \frac{i}{2 \kappa_0} \int dt \ 4 \dot{F}^2 + i \lambda N F_0 \\
& = & \frac{ 2 i \gamma}{3 \kappa_0} F_0^3 + i \lambda N F_0
\end{eqnarray}
Where,
\begin{eqnarray}
 \gamma = \frac{K}{L_y^2 L_x \xi}. 
\end{eqnarray}
By using the initial condition for $F$ [Eq. (30)], we find
\begin{eqnarray}
S_{\lambda} & = & - i \alpha \lambda^{3/2} \\
\alpha & = & \frac{2}{3}N^{3/2} \sqrt{\frac{L_y^2 L_x \xi \kappa_0}{K}}.
\end{eqnarray}
The PDF is found from Eq. (8) by utilizing the saddle point method
\begin{eqnarray}
P(Z) & = & \int d \lambda e^{-i \lambda Z - S_{\lambda}} \\
& = & \int d \lambda  e^{-i \lambda Z - i \alpha \lambda^{3/2}}
\end{eqnarray}
Let $f(\lambda) = -i \lambda Z - i \alpha \lambda^{3/2}$ and find a $\lambda_0$ such that $f$ attains its maximum and compute that value. This gives $\lambda_0 = \left( \frac{2Z}{3 \alpha}\right)^2$ and 
\begin{eqnarray}
f(\lambda_0) = - \frac{4}{27 \alpha^2} Z^3,
\end{eqnarray}
thereby giving PDF
\begin{eqnarray}
P(Z) & = & e^{- \zeta Z^3}, \\
\zeta & = & \frac{4}{27 \alpha^2}.
\end{eqnarray}
According to the definitions, for all reasonable physical situations the parameters involved in $\zeta$ are positive definite. Note that the saddle point solution justifies our assumption that $\lambda \rightarrow \infty$ corresponds to $Z \rightarrow \infty$. Eq. (40) provides the probability of finding a blob density ($Z$ normalized by $N$). Note that when the forcing vanishes ($\kappa_0 \rightarrow 0$, $\alpha \rightarrow 0$) the PDF tails vanish ($P(Z) \rightarrow 0$).
\section{Results}
We have presented first prediction of the PDF tail of blob formation. By using simplified model for the fast convective radial transport, we have found an exponential PDF tails of the form $\sim \exp\{ -\zeta n_b^3\}$ similar to what was found in Ref.~\cite{a23} for zonal flow formation in ion-temperature-gradient mode turbulence. In this section the parameter dependencies of $\zeta$ will be studied in detail. The results will be compared with a Gaussian prediction $\sim \exp\{ -\zeta n_b^2\}$.

In Figure 1 (color online) the PDF tails of blob formation as a function of blob density ($n_b$) is shown by using the parameters $\gamma = \frac{K}{L_y^2 L_x \xi} = 0.33$ (red line, dash-dotted), $\gamma = 0.66$ (blue line, solid line) and $\gamma = 1.35$ (black line, dashed line) with $\kappa_0 = 6.0$. The Gaussian distribution (green line, dotted line) is also shown for $\gamma = 0.66$ and $\kappa_0 = 6.0$. It is clearly shown that the PDF tails from the theoretical prediction recapture the experimental results shown in Ref.~\cite{a11}, where the PDF tails can be approximately fitted as $\sim \exp\{ -\zeta n_b^{\Gamma}\}$ with $\Gamma = 2.5 - 4.0$. A decrease in the parameter $\gamma$ in Eq. (34) increases the PDF tails. The predicted PDF tails deviates significantly from the Gaussian distribution.

\begin{figure}
  \includegraphics[height=.3\textheight]{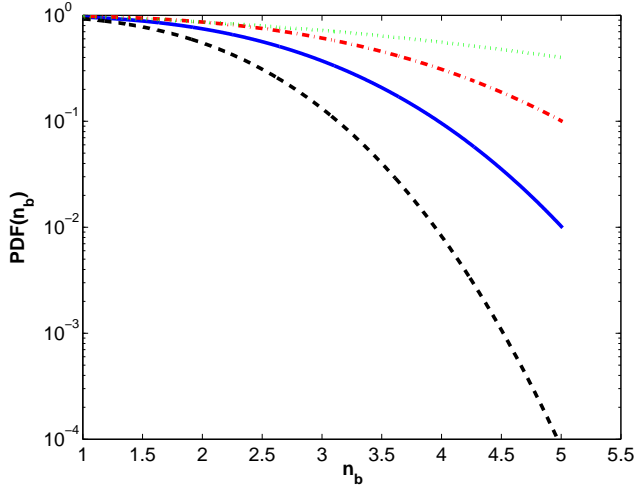}
  \caption{(Color online). The blob density PDF tail as a function of $n_b$ normalized by $N$.  The parameters are $\gamma = \frac{K}{L_y^2 L_x \xi} = 0.33$ (red line, dash-dotted), $\gamma = 0.66$ (blue line, solid line) and $\gamma = 1.35$ (black line, dashed line) with $\kappa_0 = 6.0$. the Gaussian distribution (green line, dotted line) is shown with $\gamma = 0.66$ and $\kappa_0 = 6.0$.}
\end{figure}

To elucidate the parameter dependence of the constant $\zeta$ a plot of  $1/\alpha^2$ [$\alpha$ is given by Eq. (36)] as a function of the parameter $\gamma$ with $\kappa_0 = 3.0$ (black line, dash-dotted line), $\kappa_0 = 6.0$ (blue line, solid line) and $\kappa_0 = 12.0$ (red line, dashed line) is displayed in Figure 2 (color online). Note that the PDF tails decrease as the parameter $\gamma$ decreases. In physical terms this means that increasing the cross sectional spatial scale length ($L_x$ and $L_y$) of the blob results in larger transport whereas increasing the toroidal scale length ($L_z$) decreases the PDF. Interpreting the constants in the blob speed seems to indicate that the PDF decreases for larger $v_b$. However, as shown in Sec. II the blob speed is not an independent parameter in our model but rather a combination of other parameters [see Eq. (7)]. In order to fit the theoretically predicted PDFs to experiments an estimation of the constant $1/\alpha^2$ is needed. To this end, we used experimental values in Ref.~\cite{a11} and obtained an estimate for the constant $1/\alpha^2$ as $1.5 \pm 0.5$.

\begin{figure}
  \includegraphics[height=.3\textheight]{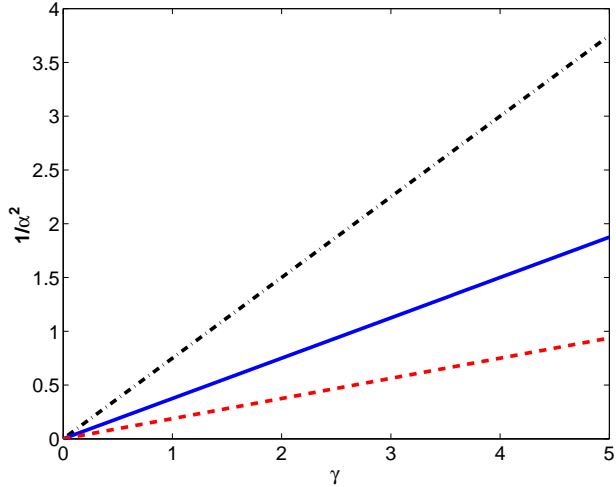}
  \caption{(Color online). $1/\alpha^2$ as a function of $\gamma$ with $\kappa_0 = 3.0$ (black line, dash-dotted line), $\kappa_0 = 6.0$ (blue line, solid line) and $\kappa_0 = 12.0$ (red line, dashed line).}
\end{figure}

\section{Discussion and conclusions}
In order to elucidate the highly intermittent turbulent transport in the edge and SOL we have considered a non-perturbative model of intermittent transport driven by blobs. By using a non-linear model of fast convective edge plasma transport derived in Ref.~\cite{a12}, we utilized the instanton calculus~\cite{a25} and~\cite{a19}-~\cite{a23} to calculate the PDF tails of blob formation. The resulting PDF tails have the exponential form $\sim \exp\{ -\zeta n_b^3\}$. Interestingly, this prediction agrees with a number of experimental results showing a highly non-Gaussian statistics of the transport at the edge~\cite{a11}. Furthermore, the PDF tails from experiments show a generic exponential form $\sim \exp\{ -\zeta n_b^{\Gamma}\}$ with $\Gamma = 2.5 - 4.0$, corroborating our prediction. Note that this is the first calculation of PDF tails of blob formation, which were shown to be strongly intermittent. Considerable transport can however be mediated by rare events of large amplitude assuming that density blobs cause radial transport.

It is also of interest to study the momentum flux driven by blobs. To this end, we replace blob formation ($\lambda \int d^2x dt n_b \delta(t)$) in Eq. (13) by the blob flux ($\lambda  N \gamma \int d^2x dt [n_b \frac{\partial }{\partial y}(\frac{1}{n_t} \frac{\partial n_b}{\partial y})] \delta(t)$) to obtain the following action
\begin{eqnarray}
S_{\lambda} & = & -i \int dt \bar{N} \left( \dot{F} + \frac{v_b}{L_x} F + K \frac{2}{L_y^2 L_x \xi} F^2 \right) \nonumber \\
& + & \frac{1}{2} \kappa_0 \int dt \bar{N}^2 \nonumber \\
& + &  2 i \lambda N \gamma \int dt F^2(t)  \delta(t).
\end{eqnarray}
Using Eq. (42) to find the instanton solutions gives us similar results as before, but with the initial condition $F_0 \propto \lambda$. This initial condition then makes the scaling for the PDF tails of momentum flux (here denoted by $Z$) as $P(Z) \sim \exp{\{-\zeta_Z Z^{3/2}\}}$. This is similar to what was found for momentum flux in drift wave turbulence~\cite{a19}-~\cite{a22}.

It is interesting to note that the exponential scalings of the predicted PDFs are similar to those found for zonal flow formation and momentum flux in ITG turbulence. The reason for this ubiquitous exponential PDFs with the same scaling is because the order of the highest non-linear interaction terms in the governing equations is the same, giving the same dependence of the large parameter $\lambda$ in the initial conditions ($F_0 = F(0)$), and thus similar exponential scalings of the PDF tails. The non-linear term can be easily seen to be quadratic in blob density $n_b$ from Eq. 4.

We have shown that the PDF tails of blob formation depends on the characteristic scale lengths ($L_x, L_y, L_z$). Although, there is no direct dependency on the blob velocity which is not an independent parameter in our model, it is indicated that the PDF tail is inversely dependent on blob speed (due to decreasing $\zeta$). Moreover the size of the blob is crucial for the PDF, with a larger poloidal ($L_y$) and larger radial extension ($L_x$) giving larger PDF tail.

We note that a non-Gaussian scaling of the PDF (the exponent of $n_b$) is found even when the forcing is Gaussian, although the exact exponent may depend on the temporal and possibly spatial correlation of the forcing ($f$). In the present paper, the forcing is chosen to be temporally delta correlated for simplicity. The source of the forcing is assumed to be the fluctuations. In general it would be desirable to identify a relation between the forcing and the turbulence amplitude. However, this is still an open problem and unfortunately outside the scope of the present paper.

Finally, we find a good agreement between our predicted PDFs and the experimental results reported in Ref.~\cite{a11}. In particular the exponential scaling is very similar to that found in experiments for reasonable parameter values. These findings strongly suggest that the experimental results are due to intermittent phenomena coming from edge turbulence. In a future publications we will consider the issues of blob generation and the asymmetrical PDFs found in many experimental and numerical work.

\section{Acknowledgment}
The authors wishes to acknowledge useful discussions with S. I. Krasheninnikov, J. Myra and F. Sattin. This research was supported by the Engineering and Physical Sciences Research Council (EPSRC) EP/D064317/1.

\end{document}